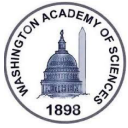

# The Transport of Extrusive Volcanic Deposits on Jezero Crater Through Paleofluvial Processes

Antonio J. Paris, Evan Davies, & Kate Morgan

Planetary Sciences, Inc.

## Abstract

Jezero, an impact crater in the Syrtis Major quadrangle of Mars, is generally thought to have amassed a large body of liquid water in its ancient past. NASA spectra of the proposed paleolake interpret the youngest surface unit as olivine-bearing minerals crystallized from magma. In early 2021, the *Perseverance* rover landed at the leading edge of a fan-delta deposit northwest of Jezero – an area argued to have experienced two distinct periods of fluvial activity. Surface imagery obtained by *Perseverance* reveal partially buried and unburied vesicular and non-vesicular rocks that appear volcanic in origin, emplaced sometime during the Noachian–Hesperian boundary. The absence of volcanic extrusive features along the fan-delta deposit, however, have made the origin of these ballast-like deposits a matter of contention among planetary scientists. To establish the origin of these basalt-like rocks, a comparison was made between analogous deposits on the Moenkopi Plateau in Arizona with similar deposits imaged by *Perseverance* on Jezero. The search for geologic analogs along the Moenkopi Plateau were guided by observable similarities in surface geomorphology, influenced and modified by fluvial, eolian, and past volcanic activity, primarily from the Late Pleistocene–Holocene boundary. By analyzing surface imagery taken by *Perseverance* and comparing it with the analogue site, we hypothesize that the exposed vesicular rocks imaged by *Perseverance* were likely transported into the paleolake by geomorphic interactions, specifically fluvial processes – similarly to the deposits that were transported along drainage patterns we observed on the Moenkopi Plateau.

## Introduction: Jezero Crater

Jezero Crater is located at latitude 18.38˚N and longitude 77.58˚E in the Syrtis Major quadrangle of Mars (United States Geological Survey, n.d.). The impact crater has a diameter of approximately 49 kilometers and was formed in the late- to mid-Noachian ~4 Gya (NASA, 2020b). Several studies propose the crater was once flooded with liquid water, being home to an ancient river delta, and has undergone significant periods of fluvial and lacustrine activity (Zastrow & Glotch, 2021). Orbital imagery obtained by NASA's High Resolution Imaging Science Experiment (HiRISE) suggests valley incisions from the paleo-river Neretva Vallis flooded Jezero from the northwest, forming a fan-delta deposit along the crater's rim (Figure 1) (Brown, et al., 2020). Although previous studies of the area have indicated two distinct periods of aqueous activity within Jezero, an analysis of impact craters along inflowing valley networks strongly suggests that fluvial activity in the region ceased by the Noachian-Hesperian boundary (Fassett, 2008).

Syrtis Major, the early Hesperian low relief shield volcano, lies to the south of Jezero. Ancient lava floods from Syrtis Major onlap the southern Noachian exposures of the Nili Fossae, the rim of the Isidis Basin, and many of the ancient fluvial networks along the Syrtis Major quadrangle adjacent to Jezero (Hiesinger & Head, 2004). A recent study proposed that large scale paleo-glaciers could have stopped lava flow from the volcano, and then diverted a network of fluvial systems draining into the Isidis Basin, including Jezero (Matherne, 2019).

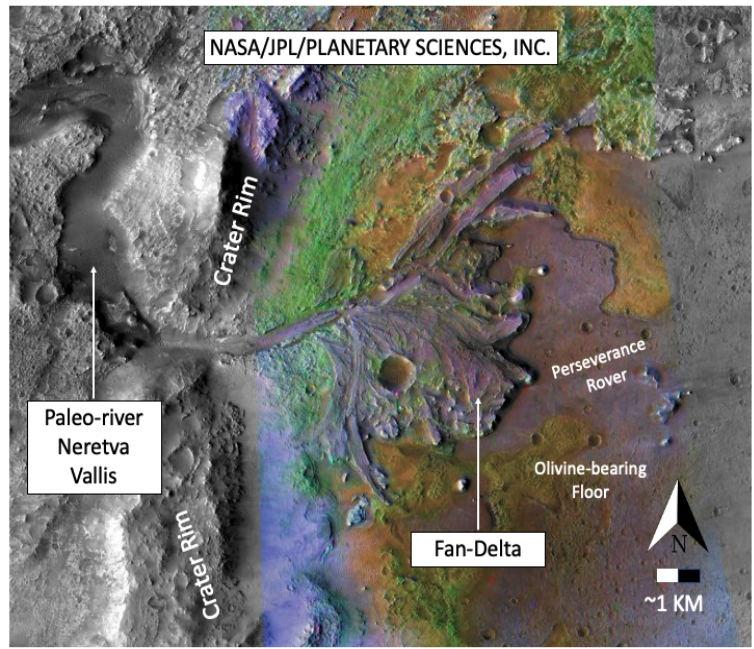

Figure 1: NASA HiRISE image of Jezero Crater and the proposed fan-delta east of the paleo-river Neretva Vallis.

**Data Collection**

**HiRISE and Context Camera (CTX)**

The NASA HiRISE and CTX imagery used in this investigation were available through NASA's Planetary Data System (PDS) and the University of Arizona's Lunar and Planetary Laboratory. Both cameras are part of the Mars Reconnaissance Orbiter (MRO) currently orbiting Mars. HiRISE can view the surface of Mars with a high-resolution capability, up to ~30 centimeters per pixel, while the CTX camera can observe at ~6 meters per pixel (University of Arizona, n.d.). The MRO also observes the Martian surface earlier in the day; thus, more geographic features are evident due to the raking sunlight along the ground surface features (Walden et al., 1998).

**Mastcam-Z (Left and Right) and Hazard Avoidance Cameras (HazCams)**

The surface imagery of Jezero used in this investigation were taken by Mastcam-Z and HazCams. The Mastcam-Z cameras are a multispectral, stereoscopic imaging instrument onboard the *Perseverance* rover. Mastcam-Z consists of two zoom cameras mounted on the remote sensing mast of the rover. The camera has a 3.6:1 zoom capable of broadband red/green/blue imaging and narrow-band visible/near-infrared color imaging. The sensor consists of an ON-Semiconductor KAI-2020CM CCD with an array size of 1600 x 1200 pixels and can image fields of view from 5° to 15° (NASA, 2020a). The HazCams, on the other hand, detect potential hazards in the front and rear pathways of the rover, including large rocks, trenches, and sand dunes. The Hazcams acquire color stereo images of the surface with a 136° to 102° at 0.46 mrad/pixel. Images taken by *Perseverance* have been made available through NASA's Jet Propulsion Laboratory.

**Thermal Emission Spectrometer (TES)**

The TES, one of five instruments on the Mars Global Surveyor spacecraft, collects two types of data: hyperspectral thermal infrared data from 6 to 50 μm, and bolometric visible-NIR at 0.3 to 2.9 μm (NASA, n.d.b). The TES instrument uses the natural harmonic vibrations of the chemical bonds in materials to determine the composition of gases, liquids, and solids on the surface of Mars. TES data used in this investigation were accessed through the Spectral Library at Arizona State University (ASU) and NASA's PDS. These datasets contain calibrated thermal IR radiance spectra and atmospheric and surface properties.

**Compact Reconnaissance Imaging Spectrometer for Mars (CRISM)**

CRISM is a visible-infrared spectrometer aboard the MRO and was accessed through the Java Mission-planning and Analysis for Remote Sensing (JMARS). The database is a geospatial information system developed by ASU's Mars Space Flight Facility to provide mission planning and data analysis tools to NASA scientists and instrument team members. CRISM maps the presence of minerals and chemicals, such as, iron and oxides, which can be chemically altered by water, as well as phyllosilicates and carbonates, which both form in the presence of water. CRISM measures visible and infrared electromagnetic radiation from 370 to 3920 nanometers (nm) with a spectral sampling of 6.55 nm/channel (NASA, n.d.a). For this investigation, datasets from CRISM's Multispectral Reduced Data Records were analyzed. These consisted of several strips of multispectral survey data mosaicked into a map tile (Murchie, 2006). CRISM spectra takes input data in units of I/F (also known as RADF, or Radiance Factor), allowing most reflectance and albedo quantities to be "unitless" quantities (University of Arizona, 2022).

**U.S. Geological Survey (USGS)**

The geological data used throughout this investigation was obtained from the USGS. Scholarly sources of information focused on the geologic history of Mars and Arizona, while indices in the National Geologic Map Database (NGMDB), e.g., a remote sensing inventory of the Jezero Crater and the Moenkopi Plateau, provided supplementary data. The geologic maps and data in the NGMDB have been standardized in accordance with the Geologic Mapping Act of 1992, section 31f(b), thus meeting widely accepted standards. Data obtained from the U.S. Geological Survey Earth Explorer imagery interface for high altitude aerial imagery also provided remote sensing inventory of the Moenkopi Plateau.

**Autel Robotics Unmanned Aerial Vehicle**

The MRO images (HiRISE and CTX) in this investigation were compared with electro-optical imagery obtained using a crewless aerial vehicle (UAV) operated by Planetary Science, Inc. flying at low altitude above the Moenkopi Plateau. The UAV utilized a powerful camera on a 3-axis stabilized gimbal that recorded video at 4k resolution up to 60 frames per second. It featured real-glass optics that captured aerial imagery at 12 megapixels from an altitude up to 800 meters and a range of 7 kilometers (Autel Robotics, n.d.).

**The Origin of Basalt Deposits on Jezero**

Multispectral imagery obtained by NASA's CRISM, a mission specifically designed to search for residues of materials that form in the presence of water, has characterized the surface mineralogy of Jezero and the proposed nearby ancient watershed as an olivine-bearing floor (Sun & Stack, 2020). Olivine is the common name for a suite of iron-magnesium silicate minerals found in many mafic igneous rocks that crystallize from magma and are generally found in the presence of water (Martel, 2003). The transportation of olivine and other basalt deposits into the Jezero by fluvial activity, though, remains a matter of contention. One hypothesis proposes the placement of an olivine-bearing lithology through resurfacing mechanisms ~3.82 Gya during the formation of the Isidis Basin and megabreccia in the Nili Fossae regional basement unit (Mustard et al., 2009).

The abundance of olivine on the crater floor of Jezero, however, is not consistent. Indices checks of CRISM, for illustration, assigned a value of 0.00930 (unitless) for olivine on the northwestern rim of the crater, whereas a value of 0.00090 (unitless) is assigned along the east - a differentiation of approximately 164%. Moreover, by superimposing a USGS/CTX mosaic of Jezero (ESP_042315_1985) with the CRSIM map tile, the composite noticeably depicts more olivine adjacent to the valley incisions of Neretva Vallis – inferring the deposition of olivine into Jezero was likely through paleo-fluvial processes (Figure 2).

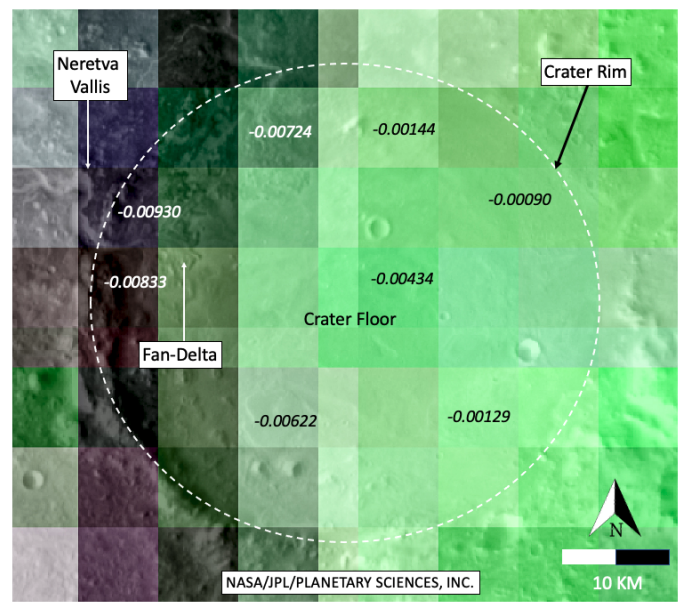

Figure 2: A composite of NASA HiRISE image of Jezero Crater and the NASA CRISM map tile with assigned olivine values.

Another hypothesis suggests that volcanic extrusions during emplacement of the Hesperian ridged plains, or a peak volcanic flux centered in the early Hesperian (which postdates the period of fluvial activity mentioned above), are responsible for basalt deposits inside the paleolakes in this quadrangle (Goldspiel & Squyres, 1991). NASA spectra (TES) of Jezero, however, infers flood basalts more than likely were transported by fluvial processed from the highlands, through Neretva Valles, and downslope into the crater. Similar to CRISM observations mentioned above, the TES map tile superimposed over the USGS/CTX image of Jezero, indicate the occurrence of basalt is greater in the northwest (51.3%) when compared to the deposits of basalt on the eastern crater floor (21.4%), a difference of approximately 82% (Figure 3).

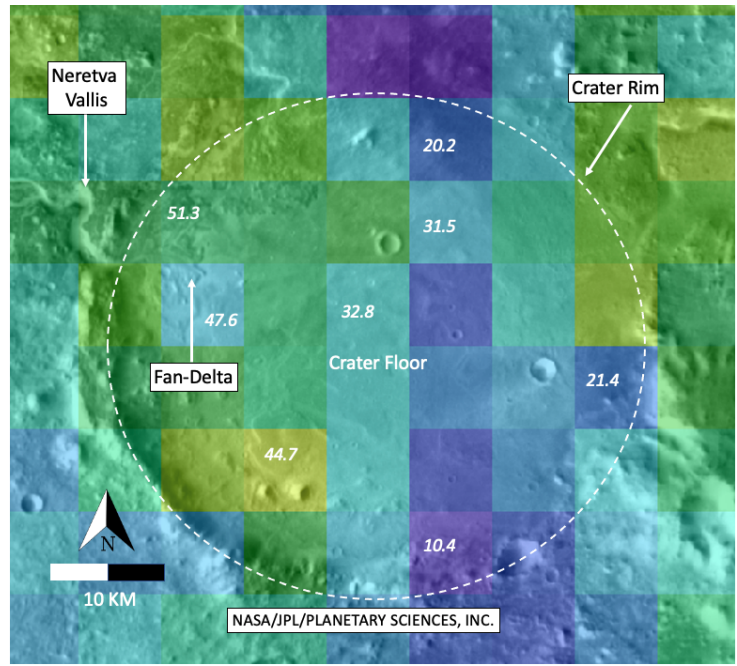

Figure 3: A composite of NASA HiRISE image of Jezero Crater
and the NASA TES map tile with assigned basalt values.

**Analog Site: The Moenkopi Plateau and the San Francisco Volcanic Field**

This investigation compares volcanic deposits on Jezero Crater with similar geologic interactions on the Moenkopi Plateau and the San Francisco Volcanic Field in northern Arizona. The Moenkopi Plateau extends from the Little Colorado River northeastward to the summit, covering 484 kilometers, with elevations ranging from 1,280 meters at the Little Colorado River to 1,700 meters on the plateau (Paris & Tognetti, 2020). The Adeii Eechii Cliffs, an erosional scarp, demarcates the southwest edge of the Moenkopi Plateau. Other major erosional scarps to the southwest include the Red Rock Cliffs and Ward Terrace (Billingsley, 1987b). Nearby, the San Francisco Volcanic Field covers approximately 2,900 kilometers, and during its ~6 Mya history has produced more than 600 volcanoes, most of which are basalt cinder cones (Priest et al., 2001).

Fluvial and eolian deposits in the analog area are dated as Holocene and Pleistocene, undivided (Billingsley, 1987a). Geomorphologically, the plateau consists of exposed sedimentary rocks and volcanic basalt deposits, with surficial deposits consisting of sand dunes, sand sheets, and landside deposits. Sedimentary rocks that consist of silica-cemented sandstone, interbedded limestone, and multi-colored shale plunge to the northeast and form northwest-trending ledges and cliffs. The bedrock in the area of study has been eroded by streams, winds, and a copious supply of loose sediment available for redeposition.

Sedimentological interactions between eolian and fluvial processes since the late Pleistocene are reflected by drainage patterns on northeasterly plunging sedimentary rocks and by the northeastward withdrawal of cuestas along the southwest boundary of the plateau. Tributary drainages, such as the Five Mile, Landmark, Tonahakaaad, Tohachi, and Gold Spring Wash, flow southwestward from the edge of the plateau toward the Little Colorado River (Figure 4). These washes, analogous to the fan-delta located on the northwestern rim of Jezero, originate in highland regions along plateaus and tend to form with a rapid change in slope from a high to a low gradient (National Geographic, n.d.). During fluvial activity, the washes transport unconsolidated sediment and rocks at a relatively high velocity due to the steep slope, leaving partially buried and unburied deposits along the leading edge of the wash (Figure 5).

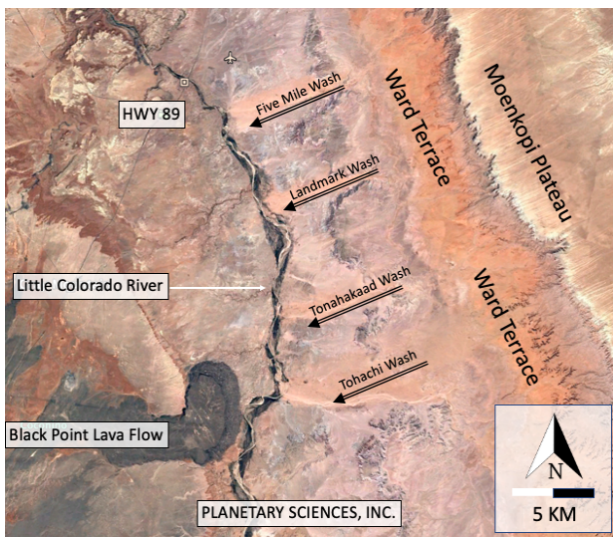
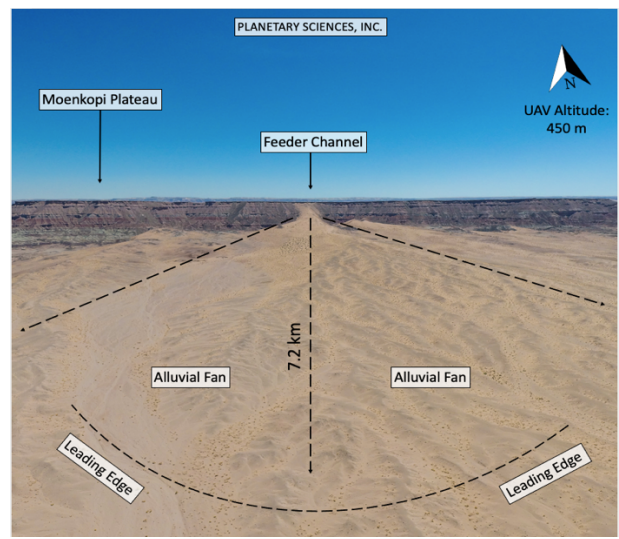

Figure 5: USGS map of the Moenkopi Plateau

Figure 5: Alluvial fan, analogous to a fan-delta, located on the Moenkopi Plateau (image taken by Planetary Sciences, Inc. UAV)

**Extrusive Basalt Deposit Analogs**

Volcanic activity has played an extensive role in the geologic development of both Earth and Mars. Many of the same magmatic processes have occurred on both planets, leading to sufficiently comparable compositions, and with similar names applied to their igneous rocks, e.g., basalt (Green & Short, 1971). Basalt, for illustration, is a volcanic rock dark in color (generally dark brown, black, or purplish-red) (King, n.d.). These rocks are somewhat low in density as a result of its numerous macroscopic ellipsoidal vesicles (holes) that formed when dissolved gases in the magma come out of solution as it erupts, forming bubbles in the molten rock, some of which remain in place as the rock cools and solidifies (Tietz & Büchner, 2018). Mafic basalt may form as part of a lava flow, typically near its surface, or as fragmental ejecta and generally consist of pyroxene and calcium-rich plagioclase feldspar. Cinder or scoria cones, such as those on the

Moenkopi Plateau and the San Francisco Peaks, violently expel lava with high gas content, and due to the vapor bubbles in this mafic lava, the extrusive basalt is formed (University of Saskatchewan, n.d).

In the following section, a sampling of Mastcam-Z and HazCam imagery of basalt-like deposits are exhibited alongside known volcaniclastic deposits unearthed on the Moenkopi Plateau. The samples, which were photographed in-situ, were found on the leading edge of fan-like topography (similar to the fan-delta on Jezero) consisting of sedimentary rocks and scattered deposits of basaltic rocks (Holocene to middle Pliocene). For the purposes of this investigation, sampling was narrowed to three specific analogs of basalt: non-vesicular, vesicular, and scoria. As a reference, the analogs on the Moenkopi Plateau were photographed alongside a field ruler with a 10 centimeter arrow facing north.

**Analog 1: Extrusive Basalt, Aphanitic, Non-Vesicular**

A mafic-like specimen on Jezero Crater was captured by Mastcam-Z (left) on Sol 86 (Figure 6a). These rocks, comparable to our analog (Figure 6b), consist of fine-grained texture with crystals too small to see with the unaided eye, have relatively low silica content and are rich in iron and magnesium. Cooling rate was rapid as heat exchanged with the atmosphere, resulting in few or no vesicles.

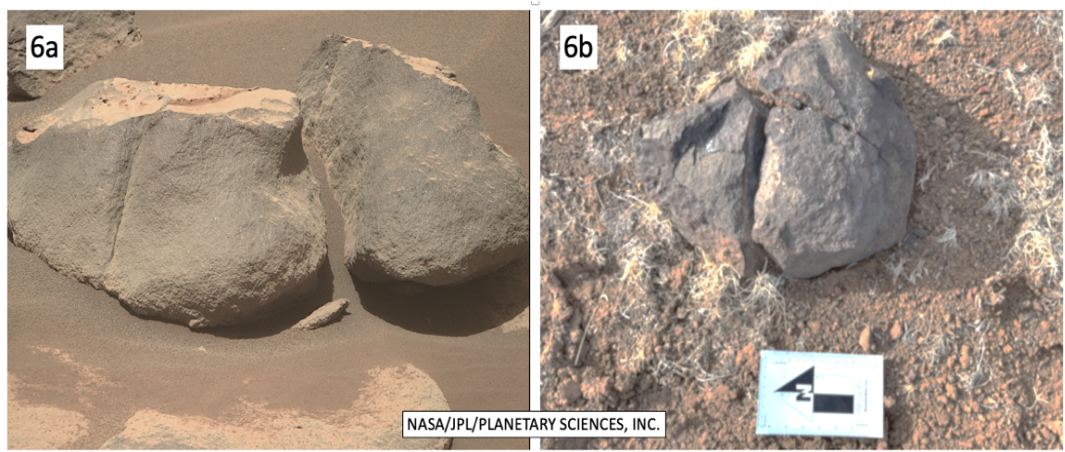

Figure 6: A Comparison of Extrusive Basalt, Aphanitic, Non-Vesicular on Jezero Crater (6a) and the Moenkopi Plateau (6b)

**Analog 2: Extrusive Basalt, Aphanitic, Vesicular:**

This specimen on Jezero Crater was taken by Mastcam-Z (left) on Sol 241 (Figure 7a). This vesicular rock is comparable to the analog on the Moenkopi Plateau (Figure 7b). Basalt often shows textural features, such as vesicles, which are formed by the expansion of bubbles of gas or steam during the solidification of the rock.

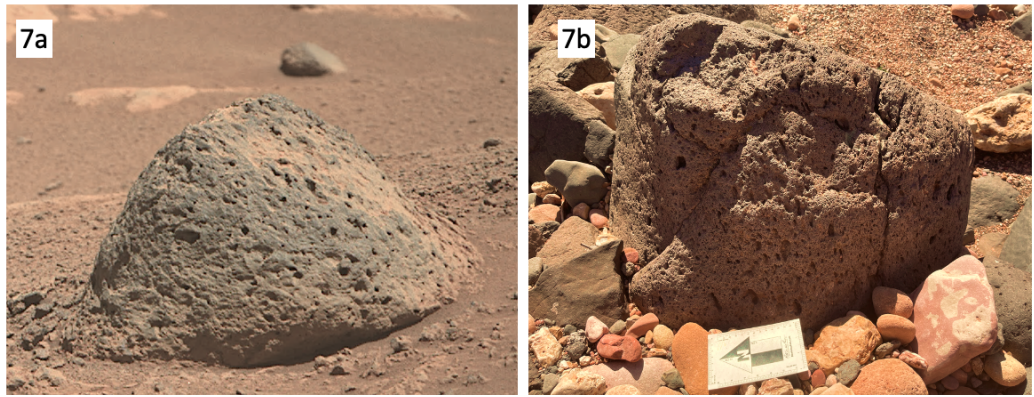

Figure 7: A Comparison of Extrusive Basalt, Aphanitic, Vesicular
on Jezero Crater (7a) and the Moenkopi Plateau (7b)

**Analog 3: Extrusive Basalt, Aphanitic, Highly Vesicular (Scoria)**

This specimen on Jezero Crater was photographed by HazCam (right) on Sol 2 (Figure 8a). Scoria, a highly vesicular type of basalt, is common along alluvial fans and washes on the Moenkopi Plateau (Figure 8b). Specimens with a rounded shape, such as these two samples, have usually been transported by fluvial processes, e.g., rivers and streams (United States Geological Survey, 2013).

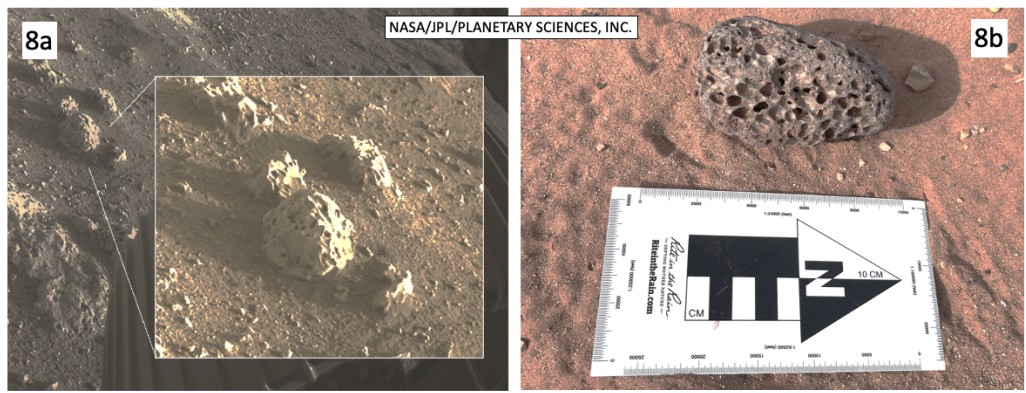

Figure 8: A Comparison of Extrusive Basalt, Aphanitic, Scoria
on Jezero Crater (8a) and the Moenkopi Plateau (8b)

## Interpretation and Conclusion

A comparison of the images taken by the *Perseverance* and the Moenkopi analogs infer comparable basaltic deposits. NASA spectra of CRISM and TES, moreover, revealed an abundance of flood basalts along the leading edge of the fan-delta (51.3%) when compared to the crater floor on the east (21.4%). We hypothesize, therefore, that the geomorphology of the Jezero

Crater region was more than likely formed during the Late Noachian, most likely in multiple, localized episodes of fluvial deposition and redeposition. During this time, deposits of flood basalts from the volcano Syrtis Major were transported east and downslope from the highlands by water, glaciation, or both along the deltas and other fluvial networks, such as the valley incisions of Neretva Vallis, into the proposed paleo-basin (Figure 9). After the evaporation of open water on the Martian surface, partially buried and unburied basalt deposits became exposed, predominantly by eolian activity. These geomorphic interactions on Jezero, specifically the paleo-fluvial processes mentioned above, are consistent with the basalt that was transported and redeposited along drainage patterns observed on the Moenkopi Plateau.

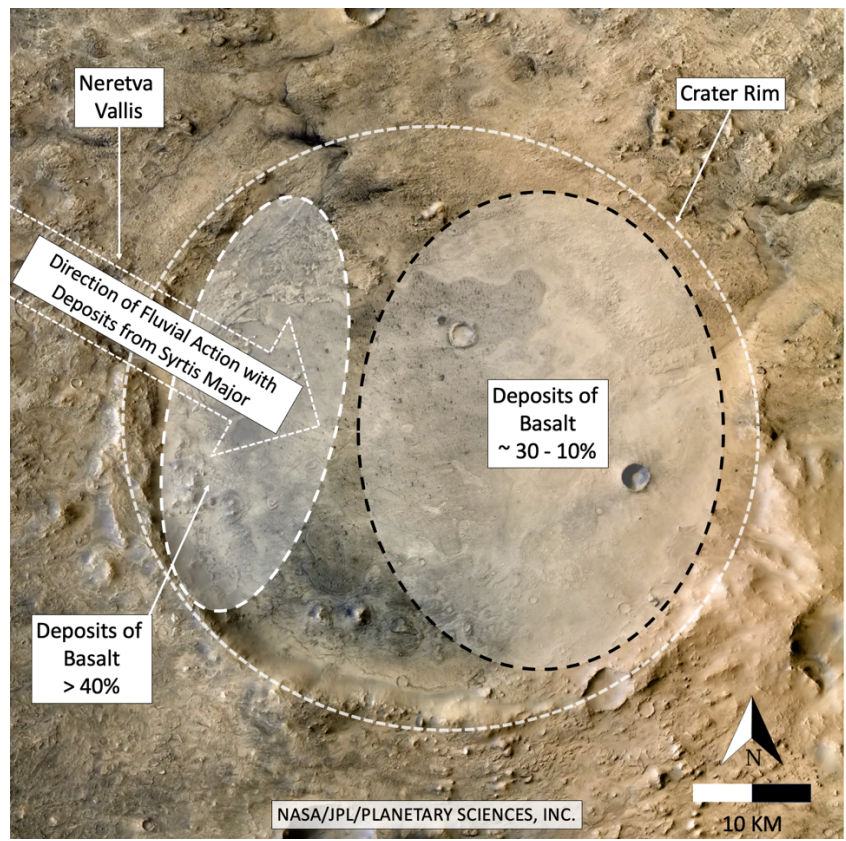

Figure 9: The transport of basalt deposits into Jezero Crater

**Biography**

Antonio Paris, Ph.D., is the Chief Scientist at Planetary Sciences, Inc., a former assistant professor of astrophysics at St. Petersburg College, Florida, and a graduate of the NASA Mars Education Program at the Mars Space Flight Center, Arizona State University. He is the author of *Mars: Your Personal 3D Journey to the Red Planet* and is a professional member of the Washington Academy of Sciences, the American Astronomical Society, and a fellow at the Explorers Club.

Evan Davies, Ph.D., is a geologist and a fellow of at the Royal Geographical Society and The Explorers Club. He is the author of *Emigrating Beyond Earth: Human Adaptation and Space Colonization* and has held a lifelong interest in planetary science, to include comets, asteroids, and impact craters.

Kate Morgan assisted Planetary Sciences, Inc. in-situ as a research assistant during this investigation. Kate collected, photographed, and made a record of the basalt deposits used for this study. She holds a Bachelor of Science degree in biology and is a recent graduate of Project POSSUM, a NASA-supported program designed to expand on bioastronautics and human factors research, providing a foundation for future space missions.

**References**

Autel Robotics. (n.d.). *EVO series*. auteldrones.com. https://auteldrones.com/collections/evo-i

Billingsley, G. H. (1987a). *Geologic map of the southwestern Moenkopi Plateau and southern Ward Terrace, Coconino County, Arizona [The Landmark, Gold Spring, Wupatki NE, Badger Spring, and Rock Head 7.5 min]*. [Map]. U.S. Geological Survey. https://www.azgs.arizona.edu/azgeobib/geologic-map-southwestern-moenkopi-plateau-and-southern-ward-terrace-coconino-county

Billingsley, G. H. (1987b). *Geology and geomorphology of the southwestern Moenkopi Plateau and southern Ward Terrace, Arizona*. U.S. Geological Survey. https://pubs.usgs.gov/bul/1672/report.pdf

Brown A. J., Viviano, C. E., & Goudge, T. A. (2020). Olivine-carbonate mineralogy of Jezero Crater. *Journal of Geophysical Research: Planets*, *125*(3), 1–30. https://doi.org/10.1029/2019JE006011

Fassett, C. I. (2008). Valley network-fed, open-basin lakes on Mars: Distribution and implications for Noachian surface and subsurface hydrology. *Icarus*, *198*(1), 37–56. https://doi.org/10.1016/j.icarus.2008.06.016

Goldspiel, J. M., & Squyres, S. W. (1991). Ancient aqueous sedimentation on Mars. *Icarus*, *89*(2), 392–410. https://doi.org/10.1016/0019-1035(91)90186-W

Green, S., & Short, N. M. (Eds.). (1971). *Volcanic landforms and surface features: A photographic atlas and glossary*. Springer.

Hiesinger, H., & Head, J. W. (2004). The Syrtis Major volcanic province, Mars: Synthesis from Mars Global Surveyor data. *Journal of Geophysical Research*, *109*(E01004), 1–37. https://doi.org/10.1029/2003JE002143

King, H. (n.d.). *Scoria*. Geology.com. https://geology.com/rocks/scoria.shtml

Martel, L. M. V. (2003, November 3). *Pretty green mineral--Pretty dry Mars?* Planetary Science Research Discoveries. http://www.psrd.hawaii.edu/Nov03/olivine.html

Matherne, C. M. (2019). *Role of glaciers in halting Syrtis Major lava flows to preserve and divert a fluvial system*. [Unpublished master's thesis]. Louisiana State University. https://digitalcommons.lsu.edu/cgi/viewcontent.cgi?article=6072&context=gradschool_theses

Murchie, S. (2006). *Mars Reconnaissance Orbiter Compact Reconnaissance Imaging Spectrometer for Mars Multispectral Reduced Data Record, MRO-M-CRISM-5-RDR-MULTISPECTRAL-V1.0*. NASA Planetary Data System. https://doi.org/10.17189/1519437

Mustard, J. F., Ehlmann, B. L., Murchie, S. L., Poulet, F., Mangold, N., Head, J. W., Bibring, J. P., & Roach, L. H. (2009). Composition, morphology, and stratigraphy of Noachian Crust around the Isidis Basin. *Journal of Geophysical Research*, *114*(2), 1–18. https://doi.org/10.1029/2009JE003349

NASA. (n.d.a). *Compact Reconnaissance Imaging Spectrometer for Mars (CRISM) Fact Sheet*. MARS Reconnaissance Orbiter. https://mars.nasa.gov/mro/mission/instruments/crism

NASA (n.d.b). *The Thermal Emission Spectrometer (TES) fact sheet*. Planetary Data System. https://atmos.nmsu.edu/data_and_services/atmospheres_data/MARS/pankine_data.html

NASA. (2020a). *Mastcam-Z for scientists*. Mars 2020 Mission: Perseverance Rover. https://mars.nasa.gov/mars2020/spacecraft/instruments/mastcam-z/for-scientists/

NASA. (2020b). *Perseverance rover's landing site: Jezero Crater*. Mars 2020 Mission: Perseverance Rover. https://mars.nasa.gov/mars2020/mission/science/landing-site/

National Geographic Society. (n.d.). *Alluvial fan*. Resource Library, Encyclopedic Entry. https://www.nationalgeographic.org/encyclopedia/alluvial-fan/

Paris, A. J., & Tognetti, L. A. (2020). Ancient river morphological features on Mars versus Arizona's Moenkopi Plateau. *Washington Academy of Sciences*, *106*(1), 59–76. https://doi.org/10.48550/arXiv.2005.00349

Priest, S., Duffield, W., Malis-Clark, K., Hendley II, J., & Stauffer, P. (2001, April 16). *The San Francisco Volcanic Field, Arizona*. U.S. Geological Survey Fact Sheet 017-01. https://pubs.usgs.gov/fs/2001/fs017-01/

Sun, V. Z. & Stack, K. M. (2020). *Geologic map of Jezero Crater and the Nili Planum region, Mars*. [Map]. U.S. Geological Survey. https://doi.org/10.3133/sim3464

Tietz, O., & Büchner, J. (2018). The origin of the term 'basalt'. *Journal of Geosciences*, *63*(4), 295–298. http://doi.org/10.3190/jgeosci.273

United States Geological Survey. (n.d.). *Map of Jezero Crater on Mars*. Gazetteer of Planetary Nomenclature. https://planetarynames.wr.usgs.gov/Page/MARS/target

United States Geological Survey. (2013). *Scoria*. Volcano Hazards Program. https://volcanoes.usgs.gov/vsc/glossary/scoria.html

University of Arizona. (n.d.). *ABOUT US: Principal & co-investigators*. Lunar & Planetary Laboratory HiRISE. https://www.uahirise.org/epo/about/

University of Arizona. (2022). *Algorithm descriptions and image processing*. UA Campus Repository. https://repository.arizona.edu/bitstream/handle/10150/641462/Global%20Photometric%20Modeling%20Algorithm.pdf?sequence=2&isAllowed=y

University of Saskatchewan. (n.d.). *Classification of igneous rocks*. Physical Geology. https://openpress.usask.ca/physicalgeology/chapter/7-3-classification-of-igneous-rocks-2/

Walden, B. E., Billings, T. L., York, C. L., Gillett, S. L., & Herbert, M. V. (1998). Utility of lava

tubes on other worlds. *Lunar and Planetary Institute Report*, 16–17. http://adsabs.harvard.edu/full/1998uisr.work...16W

Zastrow, A. M., & Glotch, T. D. (2021). Distinct carbonate lithologies in Jezero Crater, Mars. *Geophysical Research Letters*, *48*, 1–10. https://doi.org/10.1029/2020GL092365